\documentclass[sigconf]{acmart}

\usepackage{subfig}

\setcopyright{acmlicensed}
\copyrightyear{2026}
\acmYear{2026}
\acmDOI{DAC.2395}
\acmConference[DAC'26]{Design Automation Conference}{July 26-29, 2026}{Long Beach, CA}
\acmISBN{978-1-4503-2395-2/2026/07}

\begin{document}

\title{Hestia: Hyperthread-Level Scheduling for Cloud Microservices with Interference-Aware Attention}


\author{Dingyu Yang}
\affiliation{%
  \institution{Zhejiang University}\country{China}
  }
\email{yangdingyu@zju.edu.cn}

\author{Fanyong Kong}
\affiliation{%
  \institution{OKBL Technology Company Limited}\country{China}
  }
\email{fanyong.kong@okg.com}

\author{Jie Dai}
\affiliation{%
  \institution{Alibaba Group}\country{China}
  }
\email{dj378466@alibaba-inc.com}

\author{Shiyou Qian}
\affiliation{%
  \institution{Shanghai Jiao Tong University}\country{China}
  }
\email{qshiyou@sjtu.edu.cn}

\author{Shuangwei Li}
\affiliation{%
  \institution{Shanghai Jiao Tong University}\country{China}
  }
\email{shuangweili@sjtu.edu.cn}

\author{Jian Cao}
\affiliation{%
  \institution{Shanghai Jiao Tong University}\country{China}
  }
\email{cao-jian@sjtu.edu.cn}

\author{Guangtao Xue}
\affiliation{%
  \institution{Shanghai Jiao Tong University}\country{China}
  }
\email{gt\_xue@sjtu.edu.cn}

\author{Gang Chen}
\affiliation{%
  \institution{Zhejiang University}\country{China}
  }
\email{cg@zju.edu.cn}

\renewcommand{\shortauthors}{Yang et al.}

\begin{abstract}
Modern cloud servers routinely co-locate multiple latency-sensitive microservice instances to improve resource efficiency. 
However, the diversity of microservice behaviors, coupled with mutual performance interference under simultaneous multithreading (SMT), makes large-scale placement increasingly complex.
Existing interference aware schedulers and isolation techniques rely on \textbf{coarse core-level profiling} or \textbf{static resource partitioning}, leaving asymmetric hyperthread-level heterogeneity and SMT contention dynamics largely unmodeled.
We present \textbf{Hestia}, a hyperthread-level, interference-aware scheduling framework powered by self-attention. Through an extensive analysis of production traces encompassing 32,408 instances across 3,132 servers, we identify two dominant contention patterns—\textbf{sharing-core (SC)} and \textbf{sharing-socket (SS)}-and reveal strong asymmetry in their impact. Guided by these insights, Hestia incorporates (1) a self-attention-based CPU usage predictor that models SC/SS contention and hardware heterogeneity, and (2) an interference scoring model that estimates pairwise contention risks to guide scheduling decisions.
We evaluate Hestia through large-scale simulation and a real production deployment. Hestia reduces the 95th-percentile service latency by up to 80\%, lowers overall CPU consumption by 2.3\% under the same workload, and surpasses five state-of-the-art schedulers by up to 30.65\% across diverse contention scenarios.

\end{abstract}

\ccsdesc[500]{Computer systems organization~Cloud computing}

\keywords{Microservices, Inference Scheduling, Cloud computing}

\received{20 February 2007}
\received[revised]{12 March 2009}
\received[accepted]{5 June 2009}

\maketitle

\section{Introduction}
Modern cloud platforms host thousands of latency-sensitive (LS) microservices, often co-located on the same server to improve resource utilization and reduce operational costs~\cite{borg2020, Microsoft16,zhang2024ursa,stojkovic2025hardharvest}.
Such co-location inevitably introduces performance interference due to contention on shared hardware resources—including last-level cache (LLC), memory bandwidth, and interconnects~\cite{mage18,liang2023quantifying,huang2024ins}.
To mitigate such interference, production systems commonly employ \textbf{core binding or resource partitioning techniques} to isolate LS instances \cite{li2021george, preemptive17, Performance-aware20, stojkovic2025hardharvest, gajanin2025performance}.
Yet interference persists when threads from different services share the same physical core or socket, exposing the inherent limitations of coarse-grained scheduling in balancing hardware parallelism with predictable performance.

Identifying and mitigating interference in large-scale production clusters remains a \textbf{fundamentally challenging problem}.
(1) \textbf{Service Diversity}: LS microservices exhibit extreme heterogeneity in CPU intensity, memory bandwidth demands, and interference sensitivity, making unified modeling difficult.
(2) \textbf{Mutual Interference}: Contention among co-located instances is bidirectional yet highly asymmetric and nonlinear, rendering simple pairwise or additive models insufficient.
(3) \textbf{Complexity of Large-Scale Placement}: 
Selecting low-interference placements for tens of thousands of instances forms a high-dimensional NP-hard bin-packing problem~\cite{bin-packing, protean}. 
The search space expands combinatorially when accounting for hyperthread-level interactions and SC/SS contention.

Prior efforts in mitigating co-location interference can be broadly categorized as interference injection-based or trace-driven. Injection-based approaches use controlled benchmarks to emulate contention and derive sensitivity models~\cite{mars2011bubble,lo2015heracles,delim14quasar}, but incur high profiling overhead and fail to capture the diversity of real production workloads. Trace-driven methods analyze historical monitoring data to infer interference patterns~\cite{kambadur2012measuring,zhang2013cpi2,deepdive13, huang2024ins}, yet typically remain coarse-grained at the core level.
Even state-of-the-art production schedulers~\cite{mindthegap,borg2020,zhang2024ursa,luo2025embracing} that rely on Linux cpuset still operate at core granularity and overlook hyperthread sharing, while complementary SMT-focused studies~\cite{Holmes2022,liao2024retrospecting,uProcess2024} examine sibling-thread interference but lack predictive models and scalable hyperthread-level scheduling frameworks for large production clusters.


In this work, we bridge this gap by proposing \textbf{Hestia}, an attention-guided scheduling framework that operates at the hyperthread level, enabling fine-grained interference analysis and mitigation for large-scale LS microservice deployments.

Firstly, to better understand the nature of interference, we conduct an extensive analysis of trace data collected from a large-scale production cluster comprising 3,132 servers and 32,408 LS instances.
Our study identifies two dominant types of contention from an architectural perspective—\textbf{sharing-core (SC) and sharing-socket (SS) interference}—corresponding respectively to resource competition within the same physical core and within the same CPU socket.
We find that both SC and SS neighbors can substantially affect CPU utilization and service latency due to the varying degrees of hardware resource sharing.
To the best of our knowledge, this work is the first to explicitly \textbf{model interference at both core and socket levels} based on real production traces, revealing fine-grained interference behaviors that traditional core-level schedulers overlook.

Motivated by these findings, we design a hyperthread-level interference-aware scheduling framework for large-scale LS microservices.
Hestia introduces two tightly integrated components:
(1) \textbf{an attention-based CPU usage predictor}, which captures complex interference relationships among co-located LS instances by considering their SC/SS neighbors, workload characteristics, and hardware heterogeneity, effectively addressing service diversity and mutual interference; and
(2) \textbf{an interference scoring model} that leverages these predictions to guide fine-grained scheduling decisions, explicitly accounting for bidirectional and asymmetric contention.
To handle the combinatorial complexity of large-scale placement, Hestia incorporates a \textbf{topology-aware selector} that efficiently identifies candidate HT sets, reducing the search space while preserving scheduling flexibility. Together, these components enable Hestia to anticipate interference and optimize HT-level placement for heterogeneous LS services before scheduling.

We evaluate Hestia in a production cluster with 19,760 CPU cores, along with large-scale data-driven simulations.
When deployed in production, Hestia reduces 95th-percentile service latency by 21.68\%-24.6\% and lowers total cluster CPU utilization by 2.3\% under identical workloads. Hestia outperforms five representative interference-aware scheduling baselines, improving LS service performance by up to 30.65\% in our cluster.

Our main contributions can be summarized as follows:

\begin{itemize}
\item We analyze production traces from 32,408 LS instances across 3,132 servers, identifying two dominant interference types: sharing-core and sharing-socket, and systematically quantifying their impact on CPU utilization and service latency.
\item We develop a hardware-level interference model that characterizes contention among co-located LS instances, enabling accurate CPU usage prediction and guiding interference-aware scheduling decisions.
\item We propose \textbf{Hestia}, a hyperthread-level, attention-guided scheduling framework that integrates interference scoring with existing cluster schedulers to proactively mitigate contention.
\item We evaluate Hestia via deployment in a 19,760-core production cluster and large-scale simulations against five state-of-the-art baselines. Results demonstrate that Hestia reduces tail latency by up to 80\%, improves cluster-wide CPU efficiency by 2.3\%-9.75\%, and enhances LS service performance by up to 30.65\%.
\end{itemize}

\section{Background and Motivation}






\label{background}






\subsection{Microservices and Server Architecture}
Latency-sensitive (LS) microservices—such as web search, webmail, and e-commerce applications \cite{borg2020, smite, hua2023kae, zhang2024ursa}—are widely deployed in large-scale cloud platforms. They exhibit three common characteristics.
First, LS applications are composed of multiple interdependent microservices that collaboratively process user requests, forming deep call chains where performance degradation in a downstream service can propagate upstream \cite{gan2019seer}.
Second, each LS service typically runs tens to thousands of instances, load-balanced to handle massive concurrent traffic \cite{loadbalancer2020, luo2025embracing}. Any slowdown in a subset of instances can trigger heavy-tailed latency for end users.
Third, LS instances usually execute in long-lived containers—often for hours to months—making them highly sensitive to performance interference and resource contention \cite{li2021george, medea18, stojkovic2025hardharvest}.

These services execute on hierarchical multicore servers, typically with two sockets connected via Intel UPI \cite{intel}. Each socket contains multiple physical cores; each physical core exposes one or more hyperthreads (logical CPUs) that share the core’s execution resources (e.g., ALUs, FPUs) and private L1/L2 caches. Physical cores within the same socket share the last-level cache (LLC) and memory controllers, and sockets communicate over UPI. This non-uniform, layered sharing of resources—per-hyperthread, per-core, and per-socket—exposes co-located LS instances to interference at multiple architectural levels.

\begin{figure}[t]
    \centering
    \includegraphics[width=\linewidth]{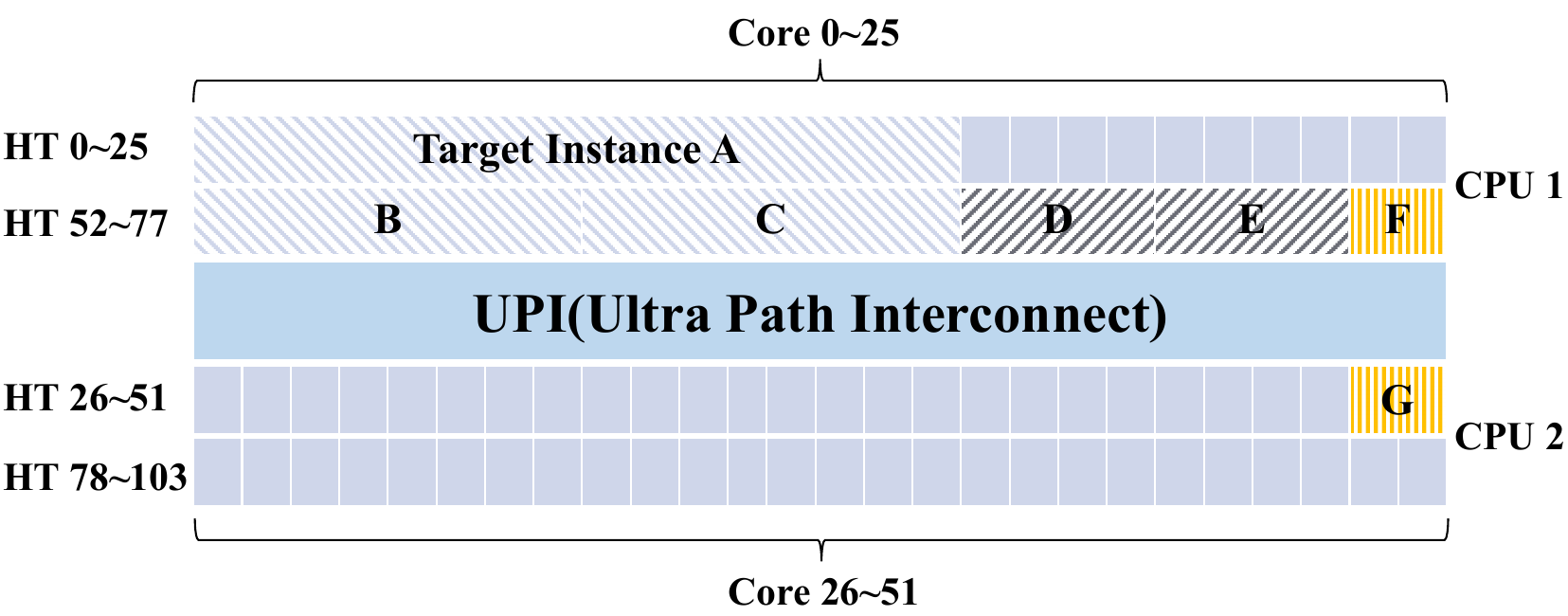}
    \vspace{-0.5cm}
    \caption{Dual-socket server with co-located LS instances.}
    \label{fig:cpu-arch}
    \vspace{-0.4cm}
\end{figure}

As illustrated in Figure~\ref{fig:cpu-arch}, interference correlates with the depth of shared resources. Instances on the same physical core (e.g., A–B or A–C) share L1/L2 and LLC resources and experience the strongest contention. Instances on different cores within a socket (e.g., D–E) compete for LLC capacity and memory bandwidth, causing moderate interference. Cross-socket pairs (e.g., F–G) share only global resources, resulting in minimal CPU interference.

Based on this hardware topology, co-located neighbors naturally fall into \textbf{sharing-core (SC)}, \textbf{sharing-socket (SS)}, and \textbf{opposite-socket (OS)} groups. This multi-level interference structure complicates performance predictability and poses a core challenge for fine-grained, interference-aware scheduling in production clusters.

\subsection{Observation and Motivation}
We analyze production traces from a 3,132-server cluster with dual-socket CPUs, focusing on 146 LS services with more than 30 instances each. This subset covers 32,408 instances and over 93\% of the total CPU quota, providing a statistically representative and meaningful view of real co-location and interference behavior.


\begin{figure}[t]
\centering
\begin{minipage}[t]{1.64in}
        \centering
         \includegraphics[scale=0.163]{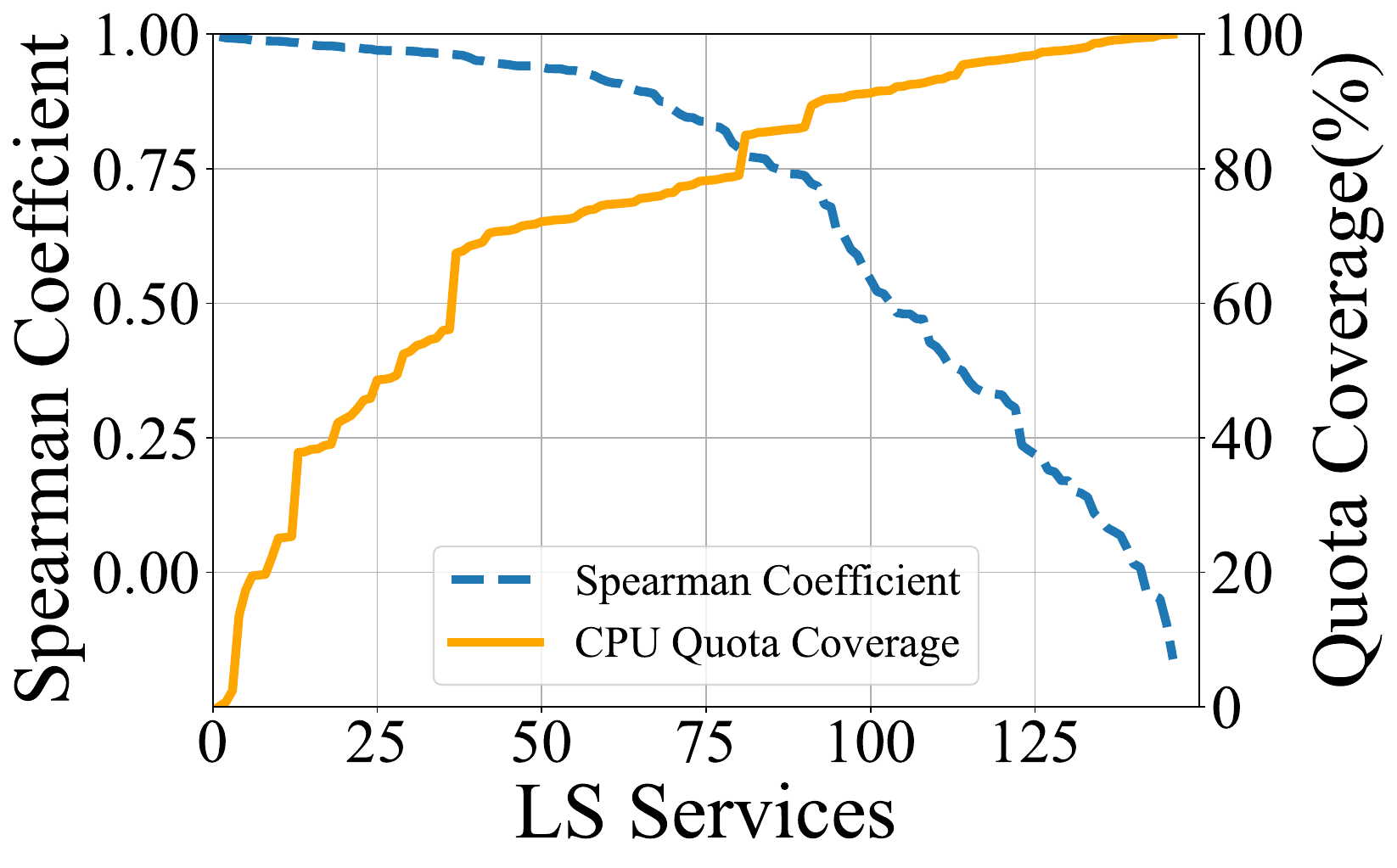}
          \caption{Spearman analysis and CPU quota coverage.}
         \label{fig:spearman}
    \end{minipage}
    \hfill
    \begin{minipage}[t]{1.53in}
         \centering
        \includegraphics[width=\linewidth]{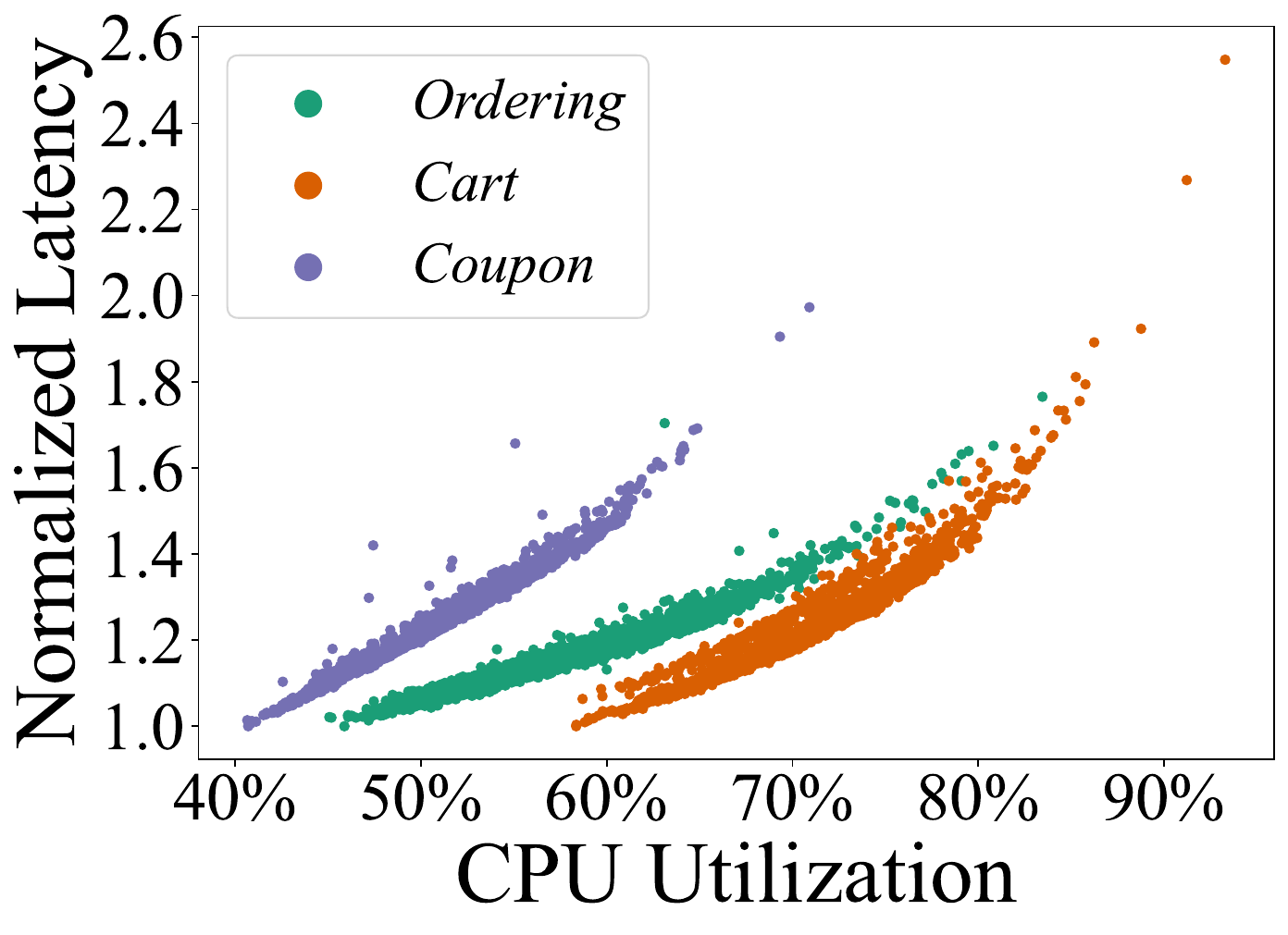}
          \caption{Correlation between CPU util. and latency.}
         \label{fig:rt-cpu}
    \end{minipage}
    \vspace{-0.4cm}
\end{figure}

\subsection{Observations}
\label{section: observation}


\begin{table*}[]
\caption{Four instance cases under distinct SC and SS interference levels.}
\vspace{-2mm}
\label{table:neighbor}
\resizebox{\linewidth}{!}{
\begin{tabular}{|c|ccccc|c|cccc|}
\hline
       & \multicolumn{5}{c|}{Target instance}                                   & \multicolumn{1}{c|}{SC neighbors}                        & \multicolumn{4}{c|}{SS neighbors} \\ \hline
       & cpu\_util.(\%) & cpi & l1d\_mpki & l1i\_mpki & l2\_mpki  & cpu\_util.(\%) & cpu\_util.(\%) & llc\_mpki & llc\_miss\_latency(ns) & memory\_bandwidth(MB/sec)    \\ \hline
case 1 &    51.50                 &   1.15  &     18.77    &   28.41      &    18.91                              &  4.61     &          7.07        &6.13 & 102.59 & 18724.86         \\ \hline
case 2 &    61.98                 &  1.36   &    21.13     &     29.66    &     20.35                                & 53.35        &              18.20   &6.72 & 104.03 & 26858.33          \\ \hline
case 3 &      62.75               &   1.42  &   20.13      &    29.80     &   20.91                                  &  27.83   &                47.63 & 8.60 & 105.72 & 48365.75       \\ \hline
case 4 &       72.51            &   1.63  &      21.91   &    31.11     &    20.93                                 &  75.60        &  50.87    & 8.07 & 107.86 & 54454.63                       \\ \hline
\end{tabular}
}
\vspace{-2mm}
\end{table*}

\noindent
\textbf{Observation 1: CPU utilization of LS instances is positively correlated with their latency.}

We define the CPU utilization of an LS instance as the ratio of CPU time to the product of elapsed time and the number of requested hyperthreads ($requests.ht$):
$$
CPU\_utilization = \frac{CPU\_time}{(requests.ht * time\_real\_passed)}
$$
Notably, high CPU utilization does not necessarily indicate efficient computation—cache misses and intra-core resource contention (e.g., between hyperthreads sharing a core) can inflate utilization without delivering proportional progress.

To quantify the relationship between utilization and performance, we compute the Spearman correlation between CPU utilization and request latency across 146 LS services. The results reveal a strong and consistent positive correlation. As shown in Figure~\ref{fig:spearman}, 78 services have correlation coefficients above 0.8, representing 78.7\% of the total CPU quota. High-traffic services—Ordering, Cart, and Coupon—display particularly clear trends, where latency rises steadily as utilization increases (Figure~\ref{fig:rt-cpu}).

These observations suggest that CPU utilization serves as a robust indicator of latency inflation. Its strong monotonic relationship with request latency enables early detection of performance degradation, making it a practical signal for proactive performance management in LS workloads.

\begin{figure}[t]
     \centering
     \captionsetup[subfloat]{captionskip=0pt, farskip=0pt}
     \begin{subfloat}[With SC neighbors]{
         \centering
         \includegraphics[width=0.46\linewidth]{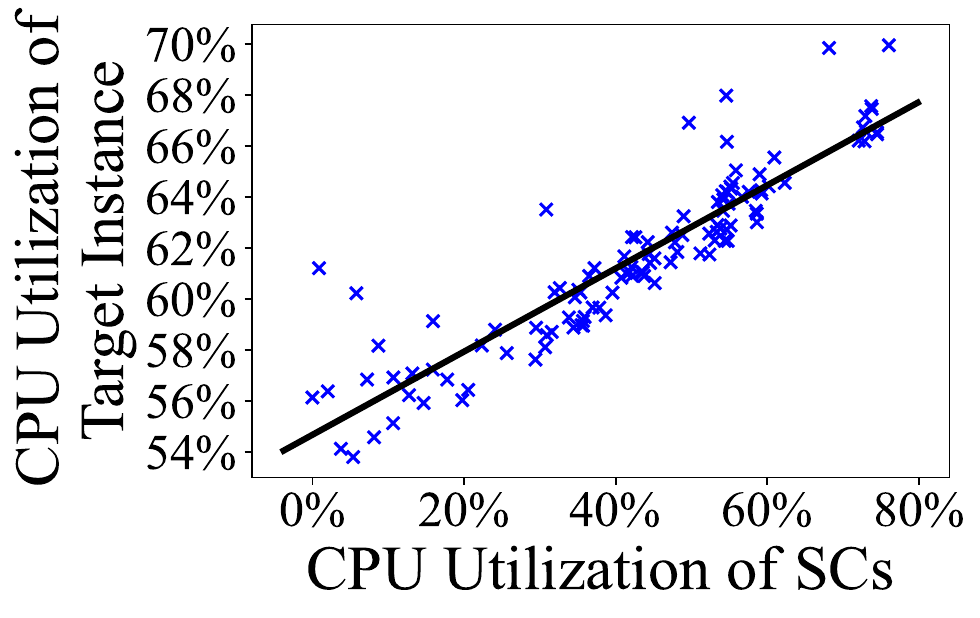}
         \label{linear-ht}
         }
     \end{subfloat}
     \hfill
     \begin{subfloat}[With SS neighbors]{
         \centering
         \includegraphics[width=0.46\linewidth]{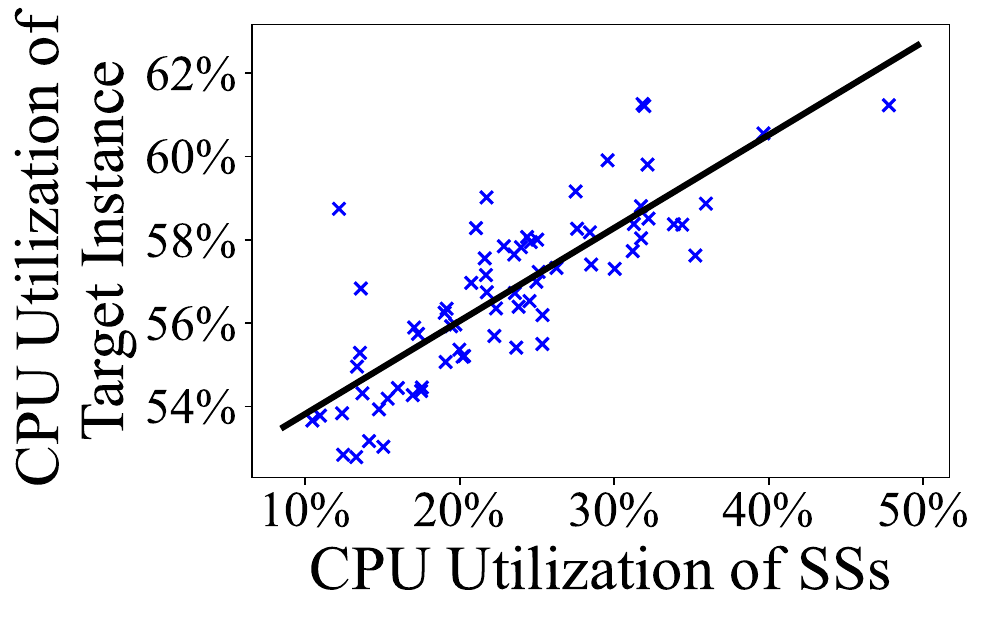}
         \label{fig:linear-socket}
         }
     \end{subfloat}
     \hfill
     \vspace{-2mm}
        \caption{Interference between $Ordering$ instances and their SC/SS neighbors.}
        \label{fig:linear}
        \vspace{-0.4cm}
\end{figure}


\noindent
\textbf{Observation 2: The CPU utilization of LS instances exhibits strong correlation with that of their SC and SS neighbors.}

We investigate whether the utilization of an LS instance is affected by the load of its co-located neighbors. Using the \textit{Ordering} service (11.70\% of total CPU quota) as a representative example, we analyze 114 instances whose SC neighbors vary in utilization while SS neighbors remain relatively stable.
As shown in Figure~\ref{linear-ht}, LS instances exhibit an almost linear utilization coupling with their SC neighbors\footnote{SC utilization denotes the average utilization of opposite hyperthreads on the same core, while SS utilization refers to the average utilization of other hyperthreads within the same socket.}.
A similar trend is observed with SS neighbors (Figure~\ref{fig:linear-socket}), whereas no clear correlation is found with opposite-socket (OS) neighbors, reflecting the stronger contention induced by intra-socket resource sharing.

To quantify this effect, we perform linear regression across all services. The average coefficient of determination ($R^2$) is 0.83, with more than 80\% of services exceeding 0.75, confirming that this coupling is both strong and consistent across diverse workloads.
This observation highlights \textbf{the importance of capturing interference from SC and SS neighbors} to enable effective interference-aware scheduling and performance diagnosis.

\noindent
\textbf{Observation 3: Performance variation arises from hierarchical resource contention among neighboring instances.}

We categorize \textit{Ordering} instances into four cases according to contention intensity from SC and SS neighbors (Table~\ref{table:neighbor}). Compared to low-interference cases, instances under high SC contention exhibit up to 23\% higher latency and 12\% more L1/L2 cache misses, whereas high SS contention leads to a 40\% increase in LLC misses. 
When both SC and SS pressures are elevated, latency rises by 78\%, CPU utilization by 40.7\%, and CPI by 41.7\%, indicating a compounded effect.

These results reveal a clear hierarchy of contention: SC neighbors primarily stress L1/L2 caches, SS neighbors inflate LLC and memory access delays, and their combination amplifies end-to-end degradation—even without core sharing. 

\subsection{Motivation}
The three observations jointly show that performance degradation in LS services is fundamentally driven by hierarchical interference among co-located instances. First, CPU utilization closely tracks latency inflation, making it a reliable proxy for interference intensity. Second, an instance’s utilization is strongly coupled with that of its SC and SS neighbors, indicating that interference propagates through shared hardware resources. Third, interference manifests hierarchically—SC neighbors pressure L1/L2 pipelines, SS neighbors inflate LLC and memory delays, and their combination compounds performance loss even under core-exclusive placement.
These insights motivate the need for \textbf{fine-grained, interference-aware orchestration} that operates at the hyperthread level. 

\section{DESIGN OF Hestia}
\label{section: design}
\subsection{Overview}

Motivated by the above observations, we present Hestia, a hyperthread-level scheduling framework designed to \textbf{quantify and mitigate CPU interference} among co-located LS instances. Hestia characterizes fine-grained interference patterns, predicts their impact on CPU utilization, and performs interference-aware placement to improve overall cluster efficiency.

\begin{figure}[t]
    \centering
    \includegraphics[width=0.9\linewidth]{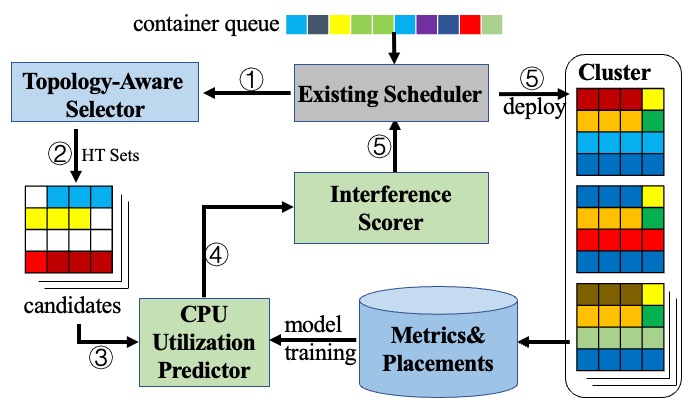}
    \vspace{-3mm}
    \caption{The overview of Hestia.}
    \label{fig:sys-arch}
    \vspace{-5mm}
\end{figure}

As illustrated in Figure~\ref{fig:sys-arch}, Hestia comprises three key components: \textbf{a Topology-Aware Selector, a CPU Utilization Predictor, and an Interference Scorer}.
(1) The \textbf{Topology-Aware Selector} identifies candidate servers and HT sets that satisfy resource and placement constraints, enabling fine-grained scheduling beyond conventional server- or core-level filtering.
(2) The \textbf{CPU Utilization Predictor} employs a self-attention–based interference model to estimate per-instance CPU usage for each candidate placement, capturing intricate SC/SS interference patterns, CPU heterogeneity, and workload dynamics.
(3) The \textbf{Interference Scorer} consolidates these predicted utilizations into a unified interference metric that directs the scheduler toward low-contention placements.

For each incoming LS instance $\tau$, Hestia filters infeasible servers, predicts utilization across candidates, and selects the placement with the lowest interference score. This workflow enables accurate interference modeling, fine-grained decisions, and seamless integration with existing production schedulers.

\begin{figure}
    \centering
    \includegraphics[width=\linewidth]{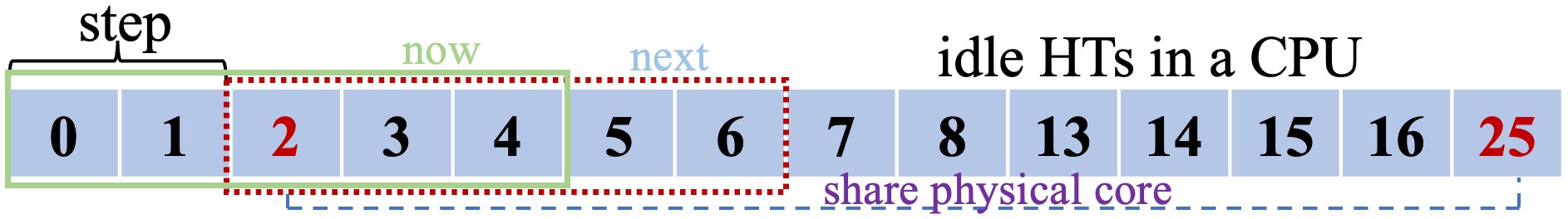}
    \vspace{-5mm}
    \caption{Sliding window for selecting HTs.}
    \label{fig:ht_window}
     \vspace{-5mm}
\end{figure}

\subsection{Topology-Aware Selector}
\label{sec: server-filter}

The \textbf{Topology-Aware Selector} is the first stage of Hestia’s scheduling pipeline, responsible for generating fine-grained placement candidates at the \textit{HT} level. Its purpose is to narrow the search space while preserving CPU topology awareness and maintaining low overhead.
The selection process consists of two steps:

(1) \textbf{Candidate Server Identification}.
Given an incoming LS instance, the selector first filters servers that satisfy resource and placement constraints. This step is fully compatible with common cluster scheduling policies—such as spread (favoring servers with more residual resources)~\cite{kubernetes,dockerswarm} and stack (packing to reduce fragmentation)~\cite{protean}—allowing Hestia to integrate with mainstream production schedulers while adding interference awareness.

(2) \textbf{Candidate HT Set Construction}.
For each feasible server, the selector enumerates valid HT sets that could host the instance. Unlike coarse-grained server- or core-level selection, this step must respect CPU topology to enable reliable interference modeling. Hestia applies two key principles:
\begin{itemize}
    \item \textbf{Socket locality}. Each instance is confined within a single socket to reduce inter-socket communication and NUMA-induced latency.
    \item \textbf{HT continuity with core isolation}. The selector chooses consecutive idle HTs on the same socket to preserve cache locality and reduce fragmentation, while avoiding HT pairs that share the same physical core unless no alternatives exist.
\end{itemize}

As shown in Figure~\ref{fig:ht_window}, Hestia employs a sliding-window mechanism to enumerate possible HT sets. The window size equals the number of requested HTs, while the step size determines the number of candidate HT sets generated per server. As the window slides across the socket’s idle HTs, only topology-valid and interference-safe HT sets are retained. 

\begin{figure*}[t]
    \centering
    \resizebox{\linewidth}{!}{
    \includegraphics[width=0.9\linewidth]{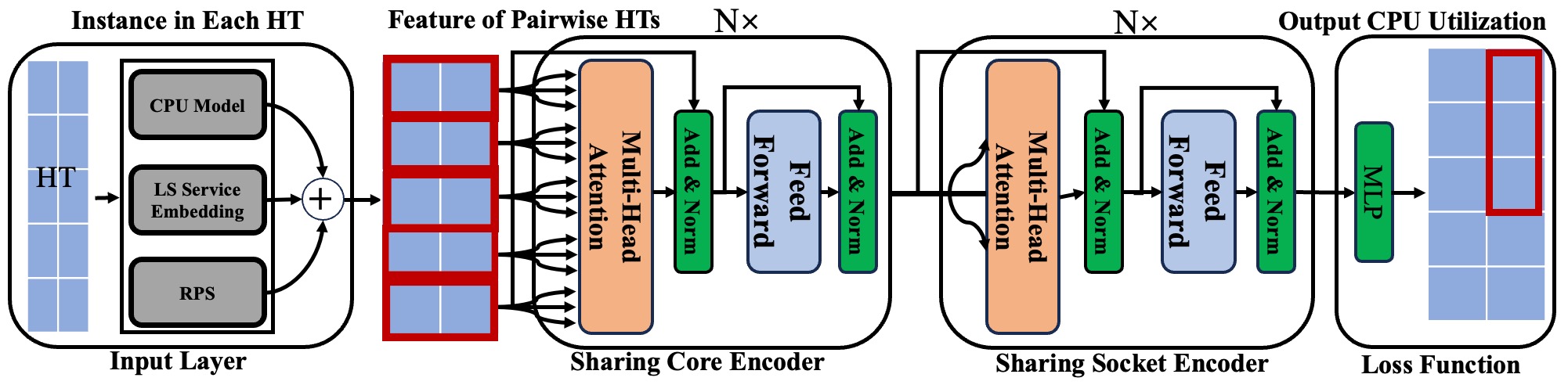}
    }
    \vspace{-3mm}
    \caption{The attention-based CPU utilization prediction model.}
    \label{fig:attention}
    \vspace{-3mm}
\end{figure*}

\subsection{Attention-guided Prediction Model}
\label{section: prediction}

Predicting CPU utilization under co-location is challenging due to \textbf{hierarchical interference} across SC and SS resources, compounded by variations in workload intensity and CPU heterogeneity. 
Traditional models (e.g., Paragon \cite{delimitrou2013paragon}, FECBench \cite{fecbench}) neglect HT binding and thus fail to capture fine-grained interference patterns present in modern SMT-enabled processors.
To overcome these limitations, Hestia adopts a data-driven, topology-aware prediction model built on a self-attention mechanism.
The attention layers dynamically learn the relative influence of neighboring instances, enabling the model to capture:
1) dynamic dependency strength among co-located services;
2) context-aware interference under varying loads and resource contention;
3) long-range cross-core influence across sockets.
As illustrated in Figure~\ref{fig:attention}, the overall architecture of the CPU utilization predictor comprises four components: \textbf{Input Layer, SC Encoder, SS Encoder, and Loss Function}.

\textbf{Input Layer}.
Each LS instance is encoded as a feature vector that captures service identity, request rate (RPS), and CPU model, forming the input tensor $I = [X_{(h, n)}W_{(n, e)}, rps, cpu\_model]$.
Here, $X_{(h, n)}$ is a one-hot encoding of microservices for the associated HT sets, where $h$ indexes the HTs allocated to the instance and $n$ indexes the microservice identity. 
$W_{(n, e)}$ is a learnable embedding matrix mapping each service to an $e$-dimensional continuous vector space~\cite{mikolov2013efficient_word_embedding}. This representation allows the model to capture both categorical service information and HT-level placement context while incorporating dynamic workload and CPU heterogeneity.

\textbf{SC Encoder}.
To model contention on shared physical cores (e.g., L1/L2 cache), this encoder applies self-attention over each HT pair sharing a core from $I$ by: $E_j = Attention(I_{(j_1, j_2)})$, where 
$I_{(j_1, j_2)}$ represents the input of two instances accommodated on core $j$.
$Attention(X)$ refers to the standard formulation in \cite{attention-is-all-your-need}, which can be formally defined as:
$
Attention(Q, K, V)=softmax(\frac{QK^T}{\sqrt{d_k}})V
$
, with $Q = W_QX$, $K=W_KX$, and $V=W_VX$.
The results across all $l$ cores are concatenated as $SC\_E = [E_1, E_2, ... ,E_l]$ and then transformed through a position-wise feed-forward network to produce the sharing-core representation $SC_H=FFN(SC\_E)$.

\textbf{SS Encoder}.
To capture LLC and memory-bandwidth contention, the SS encoder performs another attention layer on $SC_H$, producing the socket-level representation: $SS_H=Attention(SC\_H)$. 
This hierarchical design—HT-pair attention at the core level followed by attention over core-aggregated features—mirrors the layered structure of resource sharing, allowing the model to represent contention propagation from HTs to cores, and finally to the socket.

\textbf{Loss Function}.
A lightweight multilayer perceptron maps 
$SS_H$ to per-HT CPU utilization. Formally, for each hyperthread $j$, the predicted CPU utilization is $CPU_j = MLP(SS_H)_j$. For an instance $i$ that occupies a set of 
$m$ hyperthreads $H_j$, its instance-level predicted CPU utilization is computed as the average of its per-HT predictions:$CPU\_Pred_i = \frac{1}{m}\sum_{j=1}^m CPU_j$.

The model is optimized using instance-level mean squared error (MSE). Given 
$k$ instances within the socket, the loss is:$loss = \frac{1}{k}\sum_{i=1}^k(CPU\_Pred_i - CPU\_Truth_i)^2$,
where $CPU\_Truth_i$ denotes the ground-truth CPU utilization of instance $i$.
This formulation ensures that prediction accuracy is supervised at the instance level while still preserving fine-grained HT-level modeling learned by the MLP.

\subsection{Interference Scoring Mechanism}
\label{section: scoring}

Given a new LS instance and its candidate HT sets on each server, the scheduler must estimate potential performance interference and select the placement with minimal impact.
Hestia achieves this through a two-step interference scoring mechanism that quantitatively evaluates both individual and socket-level interference before deployment.

First, the baseline CPU utilization of each instance $i$ under non-interference conditions, denoted as $CPU_i^{woi}$, is derived from the utilization predictor excluding SS and SC interference.
When instances are co-located within a socket, the predictor then estimates their utilization under interference $CPU_i^{wi}$.
The relative increase in CPU utilization quantifies the interference intensity for each instance $i$:
$Score_{i} = \frac{CPU_i^{wi} - CPU_i^{woi}}{CPU_i^{woi}}$.
This normalized score captures the sensitivity of each workload to shared resource contention.

Then, to evaluate socket-level contention, Hestia aggregates individual interference scores weighted by the proportion of HTs allocated to each instance $i$: $Score_{skt} = \sum w_i *  Score_{i}$, where $w_i$ is the ratio of HTs assigned to the instance $i$ relative to all LS instances within the socket.


The aggregated socket-level interference score provides a unified metric for evaluating candidate HT placements. By ranking sockets using this score, Hestia selects the placement with the lowest expected contention, enabling fine-grained, interference-aware scheduling with minimal runtime overhead.

\section{EVALUATION}

We first conduct large-scale trace-driven simulations based on real monitoring data from our production cluster, and subsequently deploy Hestia in a controlled cluster environment to validate its performance and generalizability across representative baselines.

\subsection{Experimental setup}
Our workload is derived from production monitoring traces encompassing 32,408 LS instances across 146 services. All schedulers run on identical workload inputs for fairness. The traces span 3,132 servers equipped with a representative mix of Intel Xeon 8200/8100 series CPUs under NUMA configurations, capturing production-scale CPU heterogeneity.

\noindent
\textbf{Baselines.}
We compare Hestia with five representative schedulers, differing in their HT-binding and interference-handling strategies. All employ the spread policy for candidate server selection.
\begin{itemize}
    \item \textbf{First-Fit (FF)}: Sequentially assigns HTs based on ID order, commonly used in production clusters.
    \item \textbf{Socket-Spread}: Distributes HTs evenly across sockets, prioritizing sockets with more idle HTs.
    \item \textbf{Paragon}~\cite{delimitrou2013paragon}: Uses collaborative filtering to classify workloads and perform interference-aware greedy placement.
    \item \textbf{Kambadur}~\cite{kambadur2012measuring}: Detects “positive-beyond-noisy-interferer” cores using CPI metrics to avoid interference-prone HTs.
    \item \textbf{RCPU}~\cite{liao2024retrospecting}: An SMT-aware data-driven scheduler that reserves CPU capacity to prevent overcommitment.

\end{itemize}

\noindent
\textbf{Evaluation Metrics.}
We assess each scheduler using the following metrics:
1) \textbf{Cluster-wide efficiency}, defined as the ratio between total CPU usage and performance loss;
2) \textbf{Service performance degradation}, measured by the relative increase in CPU utilization and latency due to co-location;
3) \textbf{Average CPU utilization prediction error}, reflecting the accuracy of interference estimation.

\subsection{Overall performance evaluation}

\noindent
\textbf{Cluster Level.}
We evaluate cluster-level efficiency by computing the total CPU cores consumed to complete identical workloads under different schedulers.
Hestia achieves the lowest total CPU usage (76,420 cores), corresponding to a 9.75\% reduction compared with the production first-fit (FF) policy. In contrast, Socket-Spread, Paragon, Kambadur, and RCPU reduce CPU usage by only 0.42\%, 5.08\%, 2.65\%, and 5.21\%, respectively. 
These results show Hestia consistently improves cluster efficiency by mitigating interference among co-located LS instances.
Its performance gains stem from its neighbor-aware placement that accounts for both SC and SS interference, its fine-grained, quantitative modeling of multi-level contention across cores and sockets, and its HT-level scheduling that enables precise interference-aware placement, in contrast to baseline schedulers that operate only at coarse server granularity.

\begin{table}[t]
    \centering
     \caption{Total CPU cores required to complete identical workloads under different schedulers (lower is better).}
    \vspace{-1mm}
    \begin{tabular}{ccc}
    \toprule
        Scheduler & $total\_CPU\_Cores$↓& $CPU\_Core\_Reduction(\%)$↑\\
    \midrule
       FF (Production) &84,679 &0\\
       Socket-Spread &84,322 & 0.42\%\\
       Paragon &80,374 &5.08\%\\
       Kambadur&82,434 &2.65\%\\
       RCPU &80,264 &5.21\%\\
       \textbf{Hestia} &\textbf{76,420} & \textbf{9.75\%} \\
    \bottomrule
    \end{tabular}
    \label{tab:overall-perf}
     \vspace{-0.4cm}
\end{table}

\begin{figure*}[t]
    \centering
    \resizebox{\linewidth}{!}{
    \includegraphics[width=0.9\linewidth]{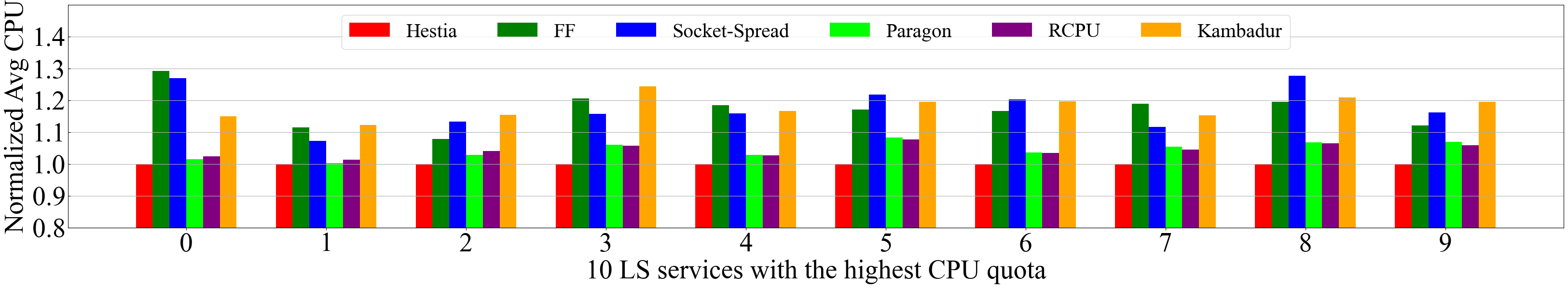}
    }
    \vspace{-5mm}
    \caption{Average CPU utilization increase of top-10 LS services (by CPU quota) over Hestia.}
    \label{fig: service-inc-overall}
    \vspace{-3mm}
\end{figure*}

\noindent
\textbf{Service Level.}
We further analyze Hestia’s impact across all LS services. Hestia consistently reduces average CPU utilization compared with five baselines. It also achieves substantial tail-latency benefits: the 95th-percentile (P95) latency is reduced by 10\%–80\% across services relative to each baseline, demonstrating Hestia’s ability to mitigate interference and improve end-to-end service performance.
Figure~\ref{fig: service-inc-overall} highlights the CPU utilization increase for the 10 LS services with the highest CPU quotas, which together account for 85.85\% of the cluster’s total CPU demand. Compared to Hestia, these baselines incur additional CPU overheads ranging from 7.91\%–29.19\% (FF), 7.16\%–27.68\% (Socket-Spread), 0.27\%–8.32\% (Paragon), 1.39\%–7.69\% (RCPU), and 12.23\%–24.36\% (Kambadur), confirming that Hestia significantly improves per-service efficiency.

\subsection{Evaluation in a Production Cluster}
We deploy Hestia on a production cluster subset consisting of 190 servers with 19,760 cores, hosting 2,009 instances across 247 LS services. 
Since global rescheduling is infeasible in our production environment, Hestia is applied only to partially affected instances, i.e., those whose CPU utilization increases by more than 20\% relative to their normal level and are thus identified as severe interference victims.
Even under this constrained deployment, rescheduling these instances with Hestia yields substantial performance improvements. 
At the \textbf{cluster level}, total CPU consumption decreases by 2.27\% under the same workload, demonstrating that Hestia alleviates interference and enhances cluster-wide efficiency even with partial adjustments.
For \textbf{service} $Ordering$, average CPU utilization drops from 81.84\% to 56.75\% (a 30.65\% reduction), and their average latency decreases from 239 ms to 180 ms (24.6\% reduction). Similar benefits occur for Coupon instances, where latency is reduced by 21.68\%. These results confirm that Hestia effectively mitigates interference and improves service performance in real production settings.


\subsection{Evaluation for CPU Utilization Predictor}

We evaluate Hestia’s attention-based CPU utilization predictor against three baselines on a constant-workload dataset to reduce load variability:
1) \textbf{Avg}: Estimates CPU utilization based on historical averages, ignoring interference.
2) \textit{FECBench}~\cite{fecbench}: Employs machine learning techniques, including K-means, random forest, and decision trees, to predict application performance under interference.
3) \textit{MLP}: Predicts CPU utilization from embedding vectors of co-located instances via a multi-layer perceptron.

\begin{figure}[t]
    \centering
        \begin{minipage}[t]{0.23\textwidth}
            \centering
             \includegraphics[height=3cm]{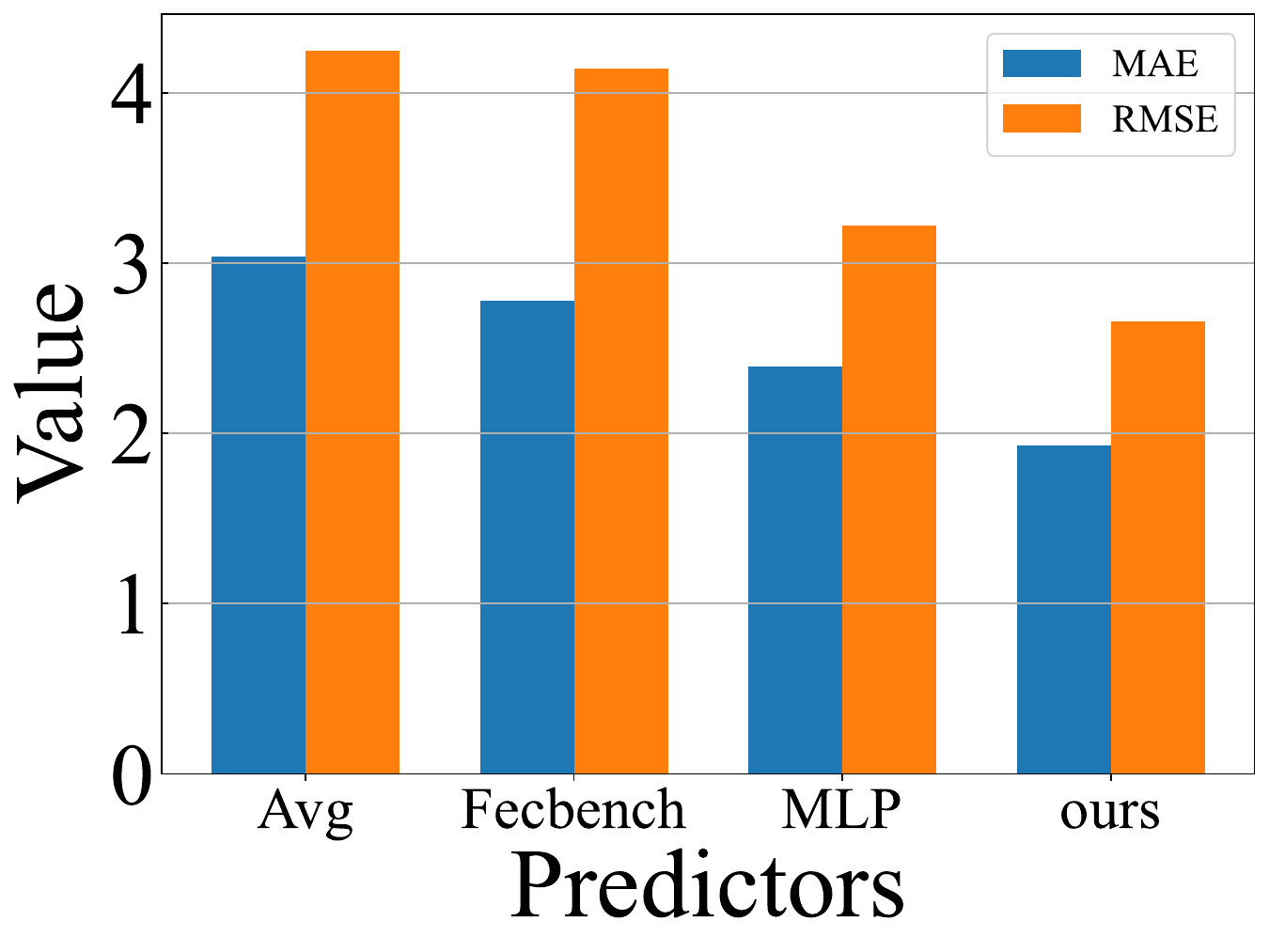}
              \caption{MAE and RMSE.}
             \label{fig: predictor-compare}
        \end{minipage}
        \hfill
        \begin{minipage}[t]{0.23\textwidth}
             \centering
            \includegraphics[height=3cm]{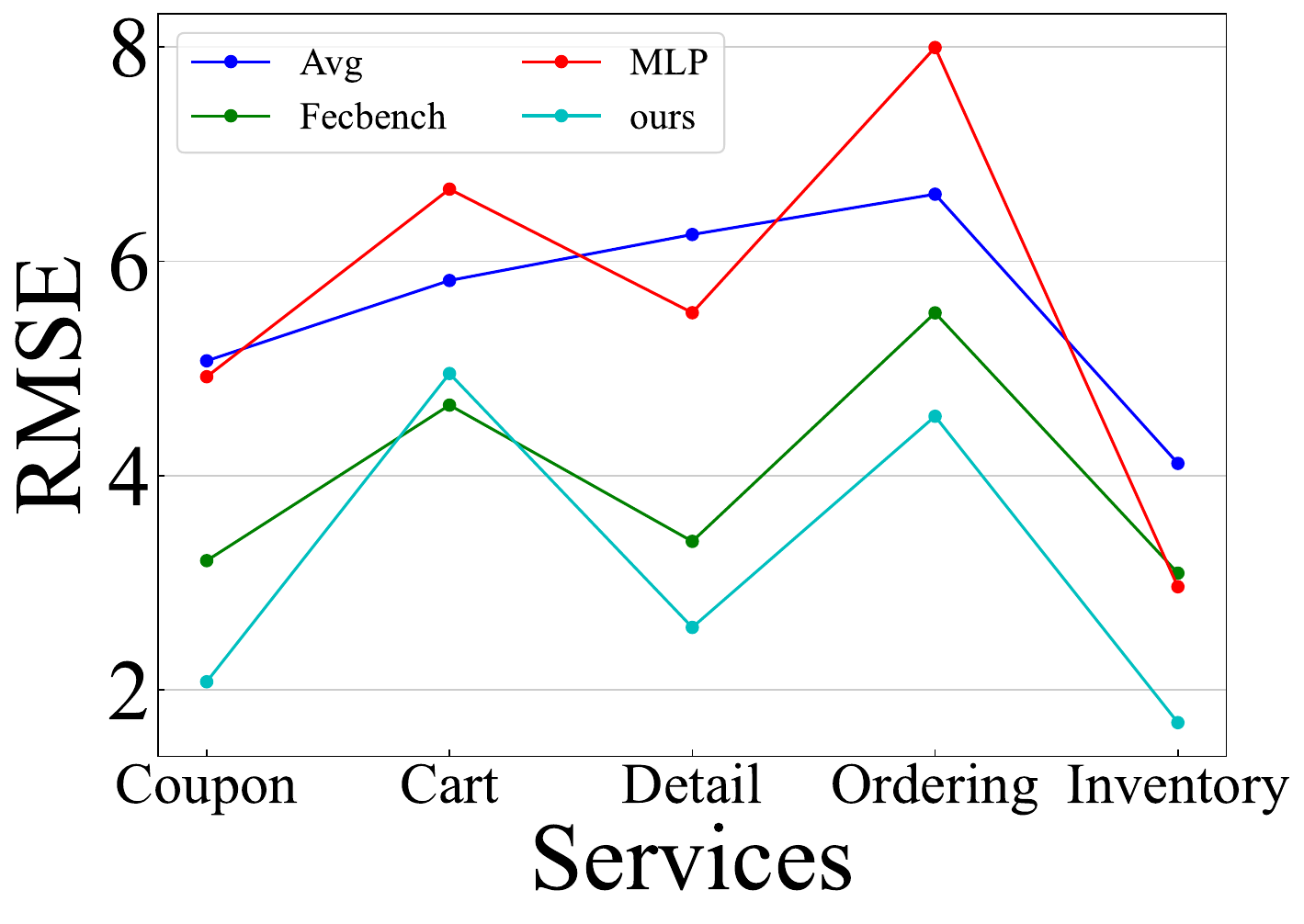}
              \caption{RMSE of services.}
             \label{fig: predictor-compare-top5-app}
        \end{minipage}
        \vspace{-2mm}
\end{figure}

Figure~\ref{fig: predictor-compare} reports the MAE and RMSE of CPU utilization predictions across all LS instances. The RMSE values are 3.04\%, 2.78\%, 2.39\%, and 1.93\% for \textit{Avg}, \textit{FECBench}, \textit{MLP}, and Hestia, respectively, while the MAE values are 4.25\%, 4.14\%, 3.22\%, and 2.66\%. 
Figure~\ref{fig: predictor-compare-top5-app} focuses on the five services with the highest CPU quotas, where Hestia outperforms the baselines for four services; for instance, it achieves 1.13\%–3\% lower RMSE for the $Coupon$ service.

Hestia’s advantage stems from explicitly modeling SC and SS interference. \textit{Avg} ignores co-located interference, while \textit{FECBench} and \textit{MLP} assume a single interference type per server, limiting accuracy. Hestia’s attention-based model captures multi-level, hierarchical interference, producing consistently more precise CPU utilization predictions.

\section{Conclusion}
This work introduces Hestia, a lightweight and non-intrusive scheduling framework that mitigates hardware-level interference for co-located LS microservices. By leveraging a self-attention-based CPU usage predictor and a fine-grained interference scoring model, Hestia proactively identifies SC/SS contention and performs hyperthread-level placement to reduce interference before it occurs. Extensive simulations and production deployment show that Hestia substantially reduces service-level latency and improves cluster-wide CPU efficiency, outperforming state-of-the-art schedulers and isolation mechanisms across diverse workloads.



\bibliographystyle{ACM-Reference-Format}
\bibliography{ref}


\end{document}